\documentclass{article}
\usepackage{graphicx, amsfonts,amsmath, tikz, tabularx, multirow, subfigure} 
\usepackage{caption, subcaption}
\usepackage[margin=2.5cm]{geometry}
\usepackage[strict]{changepage}
\usepackage{multicol}
\usepackage[affil-sl]{authblk}
\usetikzlibrary{quantikz2}
\usetikzlibrary{shapes.geometric, arrows}
\newenvironment{Figure}
  {\par\medskip\noindent\minipage{\linewidth}}
  {\endminipage\par\medskip}
  
\title{High-Entanglement Capabilities for Variational Quantum Algorithms: The Poisson Equation Case}
\author[1]{\small Fouad Ayoub}
\author[2]{\small James D. Baeder}

\affil[1,2]{Department of Aerospace Engineering, University of Maryland College Park, MD, 20742}
\affil[1]{Department of Computer Science, University of Maryland College Park, MD, 20742}

\tikzstyle{quantum} = [rectangle, rounded corners, minimum width=3cm, minimum height=1cm,text centered, draw=black, fill=red!30]
\tikzstyle{io} = [trapezium, trapezium left angle=70, trapezium right angle=110, minimum width=3cm, minimum height=1cm, text centered, draw=black, fill=blue!30]
\tikzstyle{process} = [rectangle, minimum width=3cm, minimum height=1cm, text centered, draw=black, fill=orange!30]
\tikzstyle{decision} = [diamond, minimum width=3cm, minimum height=1cm, text centered, draw=black, fill=green!30]
\tikzstyle{arrow} = [thick,->,>=stealth]

\begin{document}
\date{\vspace{-5ex}}
\maketitle

\begin{adjustwidth}{50pt}{50pt}

{\footnotesize

The discretized Poisson equation matrix (DPEM) in 1D has been shown to require an exponentially large number of terms when decomposed in the Pauli basis when solving numerical linear algebra problems on a quantum computer. Additionally, traditional ansatz for Variational Quantum Algorithms (VQAs) that are used to heuristically solve linear systems (such as the DPEM) have many parameters, making them harder to train. This research attempts to resolve these problems by utilizing the IonQ Aria quantum computer capabilities that boast all-to-all connectivity of qubits. We propose a decomposition of the DPEM that is based on 2- or 3-qubit entanglement gates and is shown to have $O(1)$ terms with respect to system size, with one term having an $O(n^2)$ circuit depth and the rest having only an $O(1)$ circuit depth (where $n$ is the number of qubits defining the system size). Additionally, we introduce the Globally-Entangling Ansatz which reduces the parameter space of the quantum ansatz while maintaining enough expressibility to find the solution. To test these new improvements, we ran numerical simulations to examine how well the VQAs performed with varying system sizes, showing that the new setup offers an improved scaling of the number of iterations required for convergence compared to Hardware-Efficient Ansatz. 

}
\end{adjustwidth}

\begin{multicols}{2}

\section{Introduction}

The Poisson equation is a partial differential equation that appears in many areas of physics and engineering, but one important application of the Poisson equation is its usage in fluid dynamics. This equation appears as a relation between the pressure and velocity field of an incompressible flow. As such, this equation holds an essential truth to any sort of analysis involving incompressible and viscous flows, for example in supernovae explosions \cite{drake2009stellar} or blood flow in arteries \cite{doost2016heart}. One of the more common classical methods of solving the Poisson is by discretizing the surface into a mesh grid and solving for the conditions at these points, reducing the problem to solving a system of linear equations \cite{finite_diff_1, finite_diff_2, finite_diff_3}. There are also some methods of solving the Poisson in recent years that include the use of modern machine-learning techniques \cite{ml_1, ml_2}. 

The focus of this paper is to develop methods of solving the one-dimensional discretized Poisson equation matrix (DPEM) on a quantum computer. This equation is one of many central pieces to the solutions of the Naiver-Stokes equation, and is key to solving a multitude of fluid-flow problems. As such, the push for greater and greater computational capabilities for simulating such fluid flows using the Poisson equation requires new techniques to be developed and improved upon. Our overall task is to solve some system of equations $A\mathbf{x} = \mathbf{b}$, where $A$ is the DPEM. There exists an algorithm known as the HHL algorithm \cite{hhl} that can explicitly solve for $\mathbf{x}$, which scales logarithmically in the system size; however, that method requires many noiseless qubits to run on reasonably sized systems. In recent literature, it has been shown that the HHL algorithm can solve the Poisson, but only for small systems \cite{small_hhl_1, small_hhl_2}. In the current era of quantum computing, we only have access to a limited number of qubits, each with some noise guarantees. This era is referred to as the Noisy Intermediate Quantum (NISQ) era of computing \cite{nisq1}. Recently, Variational Quantum Algorithms (VQAs) have had favor in the literature for near-term use, as they provide some resilience to quantum error \cite{vqa}. These VQAs use the quantum computer as the main method of computing a cost function, while a classical computer performs optimization across the parameter space of the quantum circuit.  The Variational Quantum Linear Solver (VQLS), as proposed in \cite{vqlspap}, provides a framework for which any linear system of equations can be solved. 

Typically, the VQLS is solved by representing the solution as some version of a Local-Entangling Ansatz (LEA), also known as a Hardware-Efficient Ansatz (HEA). These ansatz are used largely due to the nature of the architecture of most currently available quantum computers, such as the IBM quantum computers \cite{hea}, which can entangle qubits with their immediate neighbors using a very short native-gate circuit depth. However, the usage of ansatz involving high-entanglement schemes, what this paper introduces as Globally-Entangling Ansatz (GEA) has yet to be put into practice, and with the advent of IonQ quantum computers that boast ``all-to-all interconnectivity" of qubits \cite{ionqalltoall}, it may be wise to explore these avenues. We will show that these ansatz require much less rotation and entanglement layers than HEA, allowing for optimization algorithms to converge faster and ensuring circuit depth is reduced. The resultant scheme offers a faster convergence compared to solving the same systems with the HEA. 

Additionally, in typical analysis of solving the DPEM with VQLS, the matrix itself is decomposed into a linear combination of sub-matrices, which has a number of terms that scales either exponentially \cite{delft} or polynomially \cite{polydecomp}  with respect to system size. This means that as the size of the matrix gets larger, the number of quantum circuits required to evaluate the cost function grows exponentially or polynomially, respectively. In this paper, we propose the usage of a high-entanglement decomposition (HED), which provides a constant scaling of sub-matrices with respect to system size, containing only four terms in the final decomposition. The decomposition is inspired by the HED proposed in \cite{delft}. However, their decomposition is slightly incorrect and not optimal. Additionally, three of the terms have an $O(1)$ circuit depth, and one term has an $O(n^2)$ circuit depth. This means that when solving reasonably-sized systems, the IonQ quantum computers have the circuit depth capabilities to utilize the HED \cite{ionqaria}. The details of this result will be seen in later sections. 

There are some papers that claim that VQA schemes for solving system of equations cannot be done effectively in the NISQ era of quantum computers \cite{vqlsbad}. While their papers may hold weight for algorithms that use traditional ansatz and local entangling circuits, this paper shows that taking advantage of new types of quantum computer architectures may be the key to allowing VQAs to realize their full potential. As shown in our numerical simulations, assuming one has access to high-entanglement quantum computers allows for faster convergence to the solution and less jobs per cost evaluation. VQLS is only one of the many potential algorithms that may be benefited from the fully connected qubit assumption. This paper shows that imagining algorithms that use high-entanglement schemes may be the path forward for improving the current state of quantum computing, bringing us closer to an age of practical, commercial use of quantum computers.

\begin{figure*}
    \centering
    \begin{tikzpicture}
        \node (in1) [io] {Encode matrix $A = \sum_l c_l A_l$ and $b = \sum_l U_l$};
        \node (pro1) [quantum, below of=in1, yshift=-0.5cm] {Prepare $\ket{\psi(\boldsymbol{\theta})}$ in quantum variational ansatz};
        \node (pro2) [quantum, below of=pro1, yshift=-0.5cm] {Calculate cost function value $C$ with Hadamard Test};
        \node (dec1) [decision, below of=pro2, yshift=-1cm] {$C < \gamma$?};
        \node (pro3) [process, right of=dec1, xshift=5cm] {Find new $\boldsymbol{\theta}$ with optimizer};
        \node (pro4) [process, left of=dec1, xshift=-5cm] {$\ket{\psi(\boldsymbol{\theta})} \approx \ket{x} $};
    
        \draw [arrow] (in1) -- (pro1);
        \draw [arrow] (pro1) -- (pro2);
        \draw [arrow] (pro2) -- (dec1);
        \draw [arrow] (dec1) -- node[anchor=south] {no} (pro3);
        \draw [arrow] (dec1) -- node[anchor=south] {yes} (pro4);
        \draw [arrow] (pro3) |- (pro1);
        
    \end{tikzpicture}
    \caption{Depiction of the general workflow of the algorithm. This algorithm is based on the one proposed in \cite{vqlspap}. }
    \label{fig:workflow}
\end{figure*}
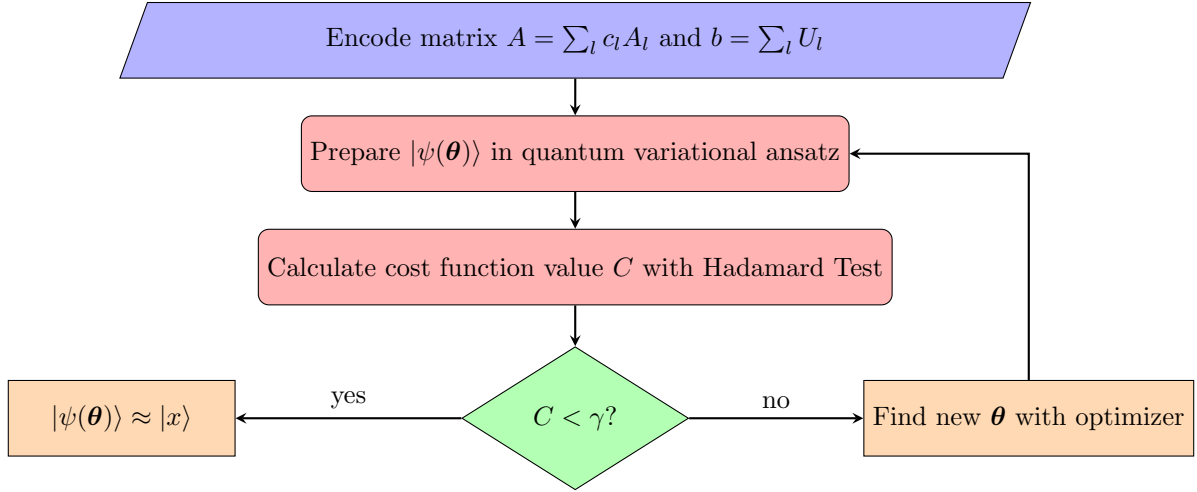

\section{Variational Quantum Linear Solver}

The following section will be rephrasing and summarizing the theory behind the VQLS \cite{vqlspap}, and will specifically describe how the VQLS structure is being implemented in this paper. 

The VQLS' goal is to solve some arbitrary linear system of equations, namely, finding $\mathbf{x}$ given some unitary matrix $A$ and some vector $\mathbf{b}$ that solves the equation $A\mathbf{x} = \mathbf{b}$. The goal of any algorithm that solves a Quantum Linear Systems Problem (QLSP) is to find some quantum circuit that prepares a state $\ket{x}$ such that $\ket{x} \propto \mathbf{x}$. 

\subsection{Algorithm Overview}

The VQLS uses a heuristic approach to solving the linear system $A\mathbf{x} = \mathbf{b}$. The schematic of the algorithm can be seen in Figure \ref{fig:workflow}. It begins by assuming that the matrix $A$ can be decomposed in some finite number of sub-matrices that can be represented as a combination and tensor product of quantum logic gates. Additionally, we assume that the vector $\mathbf{b}$ can be decomposed into a short-depth quantum circuit denoted as $U$, comprised of some sub-unitaries $U_l$, such that a quantum state $\ket{b} \propto \mathbf{b}$ can be prepared in constant circuit depth. 

\begin{align}
    A = \sum_l c_l A_l\\
    b = \sum_l U_l
\end{align}

We initially prepare some guess of the solution as a quantum state, which we call $\ket{\psi(\boldsymbol{\theta})}$, parameterized by a vector of rotations $\boldsymbol{\theta}$. This guess can be prepared using a parameterized quantum circuit known as a variational ansatz, in which the parameters $\boldsymbol{\theta}$ will determine certain rotation values of the circuit. The ansatz we use in this paper is a combination of parameterized rotation layers (using $R_y$ gates) and entanglement layers (using $CZ$ gates).

Once this guess is prepared, we evaluate our cost function (described below), and use the result to perform some type of minimization across the parameters of $\boldsymbol{\theta}$. Once the cost function has reached some convergence threshold, the optimization loop will exit, and the probability vector of the resulting guess $\ket{\psi(\boldsymbol{\theta})}$ may be read out as a proportional solution to the system. 

\begin{figure*}[ht]
    \subfigure[]{\includegraphics[width=0.43\textwidth]{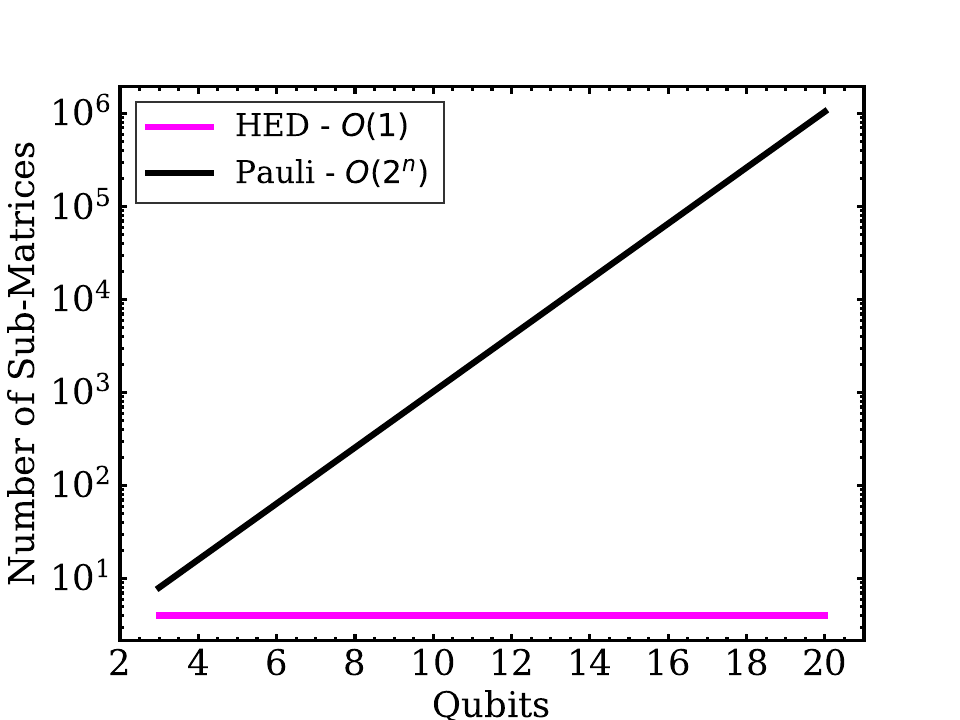}}
    \subfigure[]{\includegraphics[width=0.43\textwidth]{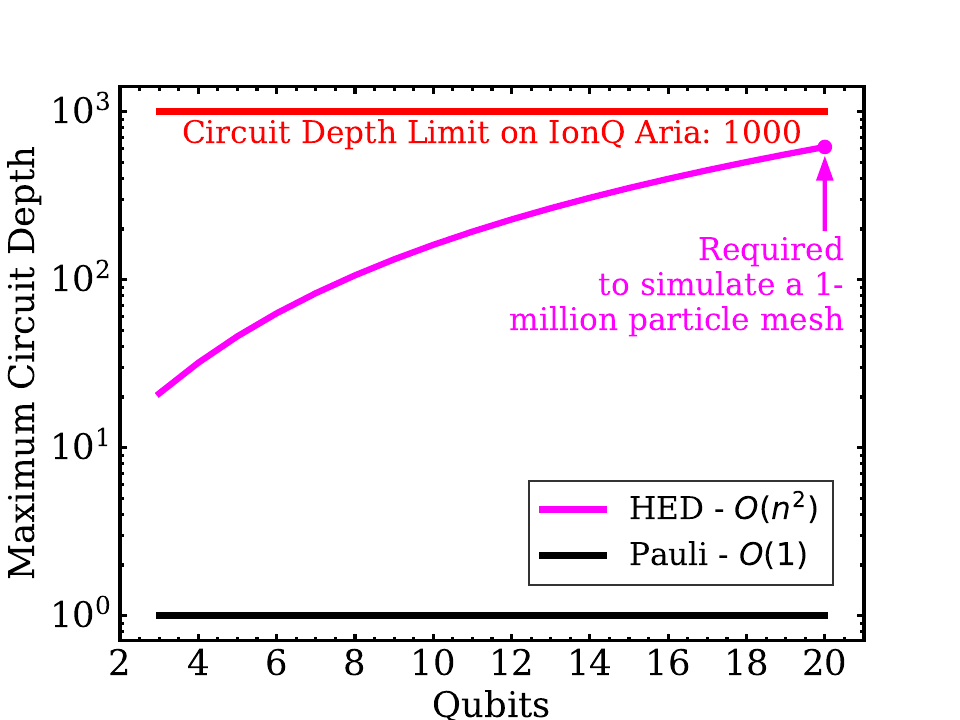}}
    \centering
    \caption{ The plot \ref{fig:HED}(a) shows the number of sub-matrices that decompose the DPEM for both the HED and Pauli decompositions (not including the circuit depth required for the ansatz or Hadamard test). It is shown that the number of sub-matrices required scales exponentially for the Pauli decomposition, but it is a constant number for the HED. \ref{fig:HED}(b) shows the maximum circuit depth that will occur when evaluating quantum circuits during the training process. While the Pauli decomposition has a constant scaling, the HED has an $O(n^2)$ scaling. However, this scaling is much more under control, as we see shown in the graph. One can already simulate a 1 million particle mesh with 20 qubits, while remaining well under the maximum circuit depth possible on the IonQ Aria machines. \cite{ionqhardware} }
    \label{fig:HED}
\end{figure*}

\subsection{The Poisson Equation}

The DPEM in 1D is as follows. It will be the matrix of interest in all of our numerical simulations. Its derivation is not the importance of this paper, but may be found in \cite{delft}.
\begin{align}
    A = \begin{bmatrix}
        2 &-1 &0 &\cdots &0\\
        -1 &2 &-1 &\cdots &\vdots\\
        0 &\ddots &\ddots &\ddots &0\\
        \vdots &\cdots &-1 &2 &-1\\
        0 &\cdots &0 &-1 &2
    \end{bmatrix}
\end{align}

The most difficult part about the VQLS algorithm is the decomposition of the desired matrix into sub-matrices that can be represented on a quantum computer. For the DPEM, there are 2 decompositions we will discuss in this paper. The first utilizes the Pauli basis of quantum logic gates, which is easy to understand and has short circuit lengths when applied to the actual algorithm. The second decomposition is more complicated and requires quantum circuits that entangle almost all qubits with each other. However, this decomposition will be favorable in later use because its decomposition has an $O(1)$ scaling of sub-matrices and a generally well-behaved scaling of circuit depth, which is $O(n^2)$. With currently available hardware (based on current circuit depth limitations of the IonQ Aria machine \cite{ionqhardware}), one could already simulate a one-million particle mesh with only twenty qubits using the HED method, as shown in Figure \ref{fig:HED}.

\subsubsection{Decomposition in Pauli Basis}

The first, and simplest decomposition of the Poisson matrix is a linear combination of multiple Pauli gates. The derivation of a recursive formula based on system size can be found in \cite{delft}, but here we list the decompositions for 4x4 and 8x8 DPEM.

N=4:
\begin{align}
    \begin{split}
        A = 2I\otimes I - I\otimes X \\
        - 0.5X\otimes X - 0.5Y\otimes Y
    \end{split}
\end{align}

N=8:
\begin{align}
    \begin{split}
        A = 2I\otimes I\otimes I - I\otimes I\otimes X\\
        - 0.5I\otimes X\otimes X - 0.25X\otimes X\otimes X\\
        - 0.25Y\otimes Y\otimes X - 0.25Y\otimes X\otimes Y\\
        - 0.5I\otimes Y\otimes Y + 0.25X\otimes Y\otimes Y
    \end{split}
\end{align}

The benefit to using the Pauli basis for decomposition is that we are left with relatively simple sub-matrices to work with when calculating expectation values of the quantum state. Also, the circuit depth is guaranteed to be $O(1)$ for each quantum circuit required to calculate the cost function. However, the Pauli basis decomposition scales exponentially with system size \cite{delft}, meaning that solving the Poisson matrix in this manner would not offer an asymptotic speedup when compared to classical methods. For this reason, the Pauli decomposition was not favored in our numerical simulations and runs on the IonQ quantum computers. 

\subsubsection{High Entanglement Decomposition}

A second decomposition of the Poisson equation exists, although it requires more complicated procedures for its generation, and it requires that the quantum computer for which it runs on can complete two or three qubit entanglement gates between almost all qubits. IonQ's latest quantum computers indeed are fully capable of such operations \cite{ionqhardware}, so their quantum computers are a prime candidate for testing this decomposition of the matrix on a real quantum computer. The HED shortens the number of expectation values required to be evaluated per iteration, has a constant number of sub-matrices, and has reasonable circuit depth for each quantum circuit required to evaluate the cost function, as will be shown in the following derivation and can be seen in Figure \ref{fig:HED}. This decomposition takes much less computational resources overall compared to the Pauli decomposition, so we use it in all our numerical simulations.  

The HED proposed in this paper is composed of only four sub-matrices. The linear combination of the matrices are as follows:

\begin{align}
    A = 2.5I - L_1 - L_2 - 0.5 L_3
\end{align}

The matrix $I$ is the identity matrix acting on all qubits. The remaining matrices are defined below.

$L_1$ is simply the $X$ gate applied to the least significant qubit.

\begin{align}
    L_1 = I^{\otimes n} \otimes X
\end{align}
This gives us the matrix

\begin{align}
        L_1 = \begin{bmatrix}
        0 &1 & & &\\
        1 &0 & & &\\
        & &\ddots & &\\
        & & &0 &1\\
        & & &1 &0
    \end{bmatrix}
\end{align}
which clearly has an $O(1)$ circuit depth. A.\ref{fig:L1}(a) shows the quantum circuit for $L_1$ for the $n=6$ case.

$L_2$ is more complicated to craft, and it requires the definition of an extra sub-matrix, $C_i$, which is composed of $CX$s and multi-$CX$s. The definition used the following notation of $CX$ gates: $CX^b_a$ describes an $X$ gate acting on $a$ controlled by $b$. $mCX$ gates simply can have more controls $b_0, b_1, \ldots$

\begin{align}
    \begin{split}
    C_i &= CX^i_{i-1} CX^i_{i-2} \cdots CX^i_{0}\\& 
    mCX^{0, 1, \ldots, i-1}_i \\& CX^i_{0} \cdots CX^i_{i-2}  CX^i_{i-1}        
    \end{split}
\end{align}

A.\ref{fig:Cgates} depicts the quantum circuits representing $C_1, C_2, \ldots, C_5$ as an example for the $n=6$ case.

$L_2$ is thus formed in the following manner
\begin{align}
    L_2 = C_1 C_2 \ldots  C_{n-1}
\end{align}
where $n$ is the number of qubits. Figure \ref{fig:Cgates}(f) shows the final result for $n=6$. This gives us the matrix
\begin{align}
    L_2 = \begin{bmatrix}
        1&  &  & & &\\
         &0 &1 & & &\\
         &1 &0 & & &\\
         &  &  &\ddots & &\\
         &  &  & &0 &1\\
         &  &  & &1 &0\\
         &  &  & &  & & 1
    \end{bmatrix}
\end{align}
The $L_2$ circuit depth is 
\begin{align}
    &3 + 5 + 7 + \ldots + 2(n-1) + 1 \\
    &= \sum_{i=1}^{n-1} 2i+1\\
    &= O(n^2)
\end{align}

Finally, we just need the $L_3$ matrix. This one is much more easily implemented. It is simply applying multi-controlled $Z$ gates and an $X$ gate on the least-significant qubit in alternating order.
\begin{align}
    L_3 = mCZ^{0, 1, \ldots n-1} \: X \: mCZ^{0, 1, \ldots n-1} \: X
\end{align}
The resultant $L_3$ matrix is 
\begin{align}
    L_3 = \begin{bmatrix}
        -1&  &       &  & \\
          &1 &       &  & \\
          &  &\ddots &  & \\
          &  &       &1 & \\
          &  &       &  & -1
    \end{bmatrix}
\end{align}
An example of the implementation of this matrix as a quantum circuit is shown for the $n=6$ case in Appendix \ref{appd:a}.1. 

\subsection{The Cost Function}
We define a cost function, which will determine how good the current guess $\ket{\psi}$ is by finding out how close $A\ket{\psi}$ is to $\ket{b}$. This cost function is inherently hybrid, in that it will be partly evaluated with a series of quantum circuits that evaluate circuit matrix arithmetic, as well as some classical matrix arithmetic. 

\subsubsection{Global Entangling Ansatz}

\begin{figure*}
    \centering
    \includegraphics[scale=0.4]{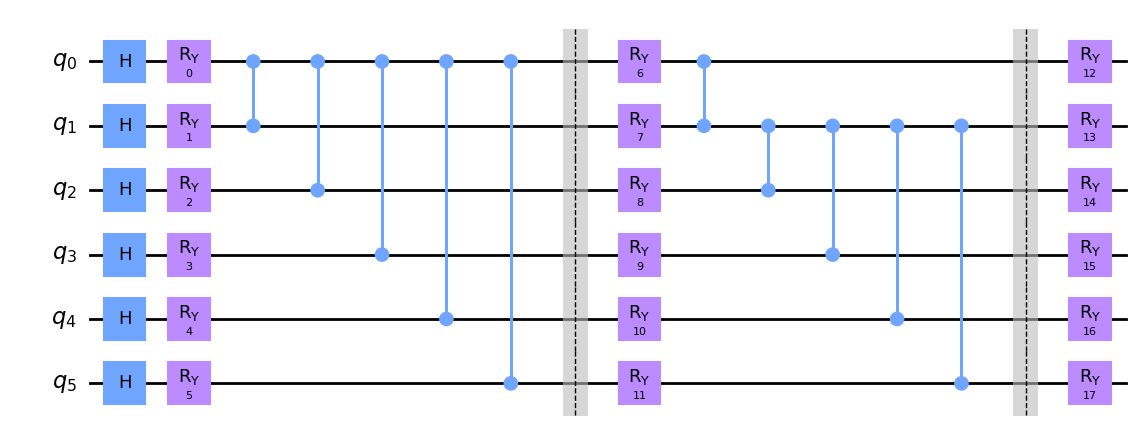}
    \caption{The global-entangling variational ansatz, defined with $3n$ parameterized gates and $CZ$ entangling gates, which prepares the $\ket{\psi(\boldsymbol{\theta})}$ state. This figure depicts the $n=6$ case. Here, we pre-condition the ansatz with the gate sequence that prepares the \ket{b} state, $U_b$; in this case, that gate sequence is $H^{\otimes n} \ket{0}$ }
    \label{fig:ansatz}
\end{figure*}

In order to generate our $\ket{\psi(\boldsymbol{\theta})}$ state, we use a Global-Entangling Ansatz (GEA), which utilizes the capabilities of the IonQ hardware. The quantum computers available through IonQ's quantum computing cloud services boast all-to-all connectivity between qubits, allowing one to easily encode two-qubit gates between any two qubits \cite{ionqalltoall}. We define the GEA with multiple layers of parameterized rotation and entanglement gates, where the parameters are defined by $\boldsymbol{\theta}$. Additionally, we pre-conditioned the ansatz with the unitary that prepares the \ket{b} state, denoted by $U_b$. We decided to see whether or not this type of entangling scheme would benefit the performance of the cost functions, and as shown through our numerical results, we found that it offered great improvements with respect to number of qubits and number of iterations when compared to the LEA (or HEA) proposed in \cite{vqlspap}. There are certain factors going into how many layers (and thus the number of parameters) the ansatz should have, but simply put, if the parameter space is too small, the solution might not exist in it at all, but if it is too large, it might take far too long to find the solution \cite{vqa}. As can be seen in Figure \ref{fig:params_vs_qubits}, the GEA has far fewer parameters than the HEA, which will make it easier to train. The high-entanglement nature of the GEA preserves its expressibility, thus making the solution to each system still well-accessible to the ansatz. The GEA used in this paper is defined with three repeated layers of rotation gates ($R_y$) and entangling gates ($CZ$), with a unique rotation parameter attributed to each $R_y$ gate. Thus, the number of parameters per variational ansatz is $3n$ where $n$ is the number of qubits defining the system size. An example of the GEA for $n=6$ can be found in Figure \ref{fig:ansatz}.

\begin{Figure}
    \centering
    \includegraphics[width=\textwidth]{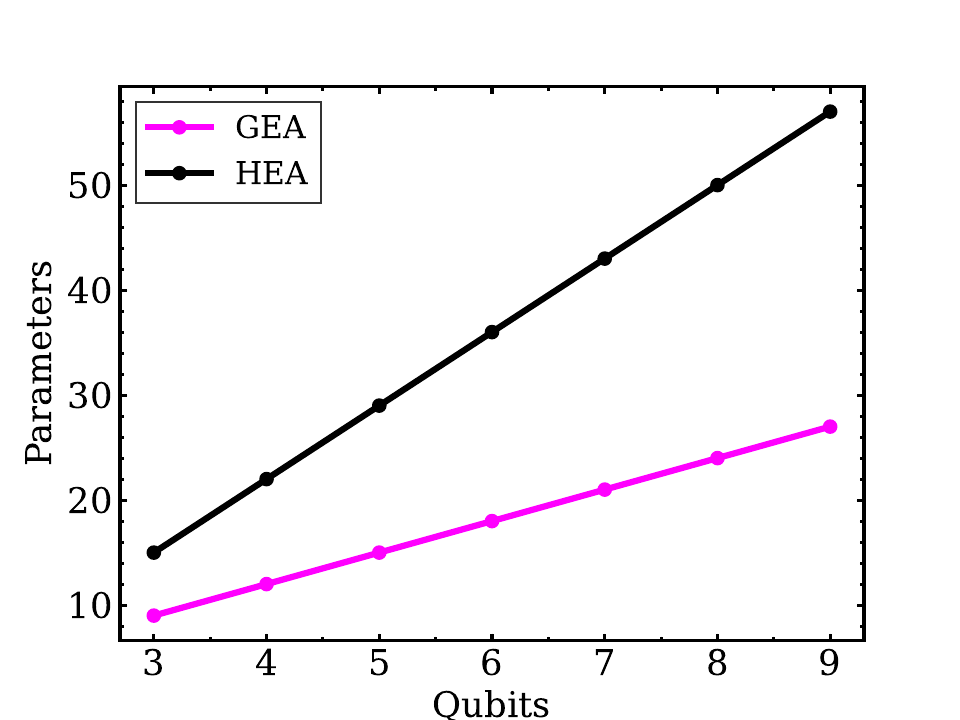}
    \captionof{figure}{Parameters versus qubits depending on the ansatz chosen. }
    \label{fig:params_vs_qubits}
\end{Figure}

\subsubsection{The Hadamard Test}

The Hadamard Test is a tool used to actually determine the expectation values of evaluating vector products of unitary matrices, or compositions of unitary matrices, with prepared states \cite{hadtest}. The general idea is that, given some initially prepared state, $\ket{\psi}$, and some unitary matrix, $U$, the value of $\bra{\psi}U\ket{\psi}$ can be determined by applying the Hadamard gate to an auxiliary qubit, then applying a $CU$ gate (a controlled version of that unitary gate) with the auxiliary qubit as the control, and finally applying another Hadamard gate on the auxiliary qubit. The resulting probabilities of measuring the auxiliary qubit as a 0 or 1 can reconstruct the expectation value as follows:

\begin{align}
    P(0) - P(1) = Re\bra{\psi}U\ket{\psi}
\end{align}

This uses a phenomenon known as ``phase kickback" and is further explained in other works \cite{vqlspap}. Since we do not use or expect any gates, matrices, or vectors to contain imaginary numbers, we neglect the calculation of the imaginary part of the expectation values. 

\subsubsection{Locally and Globally Defined Cost}

The cost function must signify how close we are to the real solution, where the larger it is, the further away we are, and the smaller it is, the closer we are. Here we inherit the ``projection" Hamiltonian from \cite{vqlspap}, defined as
\begin{align}
    H_P = \mathbb{I} - \ket{b}\bra{b},
\end{align}
so our cost function $C_G$ may be defined as the expectation value of the projection Hamiltonian, given that $\Phi = A \ket{\psi(\boldsymbol{\theta})}$:
\begin{align}
    C_G &= \bra{\Phi}H_P\ket{\Phi}\\
    &= \bra{\Phi}(\mathbb{I} - \ket{b}\bra{b})\ket{\Phi}\\
    &= \braket{\Phi}{\Phi} - \braket{\Phi}{b}\braket{b}{\Phi}
\end{align}

So when $\ket{\Phi}$ and $\ket{b}$ are orthogonal, the inner product $\braket{b}{\Phi}$ gets smaller, making the cost greater, and when the vectors get closer to each other, that inner product gets larger, making the cost smaller. However, we can normalize the cost to give it a better operational meaning \cite{vqlspap}. Taking the norm with respect to $\Phi$, we get the following new cost function:
\begin{align}
    \hat{C}_G &= \frac{\braket{\Phi}{\Phi}}{\braket{\Phi}{\Phi}} - \frac{\braket{\Phi}{b}\braket{b}{\Phi}}{\braket{\Phi}{\Phi}}\\
    &= 1 - \frac{\braket{\Phi}{b}\braket{b}{\Phi}}{\braket{\Phi}{\Phi}}\\
    &= 1 - \frac{|\braket{b}{\Phi}|^2}{\braket{\Phi}{\Phi}}
\end{align}

To break this down into quantum gate components (where $U$ is some unitary preparing the state $\ket{b}$, and $V$ is some variational unitary preparing the state $\ket{\psi(\boldsymbol{\theta})}$):
\begin{align}
    \bra{b} &= (U\ket{0})^{\dagger} = \bra{0}U^{\dagger} \\
    \ket{\Phi} &= A\ket{\psi(\boldsymbol{\theta})} = AV(\boldsymbol{\theta})\ket{0}
\end{align}

Since we know that the $A$ matrix is a linear combination of some decomposed sub-matrices, we generalize $A$ to being
\begin{align}
    A = \sum_l c_lA_l ,
\end{align}
where each $A_l$ is a sub-matrix that can be represented as a quantum gate. This gives us

\begin{align}
    \ket{\Phi} &= \sum_l c_lA_l\ket{\psi(\boldsymbol{\theta})} = \sum_l c_lA_lV(\boldsymbol{\theta})\ket{0}
\end{align}

Now we can rewrite $\hat{C}_G$ as follows:
\begin{equation}
\begin{aligned}
    &\hat{C}_G =\\
    &1 - \frac{\sum_{l, l'} c_lc_{l'}\bra{0}V(\boldsymbol{\theta})^{\dagger}A_l^{\dagger}U\ket{0}\bra{0}U^{\dagger}A_{l'}V(\boldsymbol{\theta})\ket{0}}{\sum_{l, l'} c_lc_{l'}\bra{0}V(\boldsymbol{\theta})^{\dagger}A_l^{\dagger}A_{l'}V(\boldsymbol{\theta})\ket{0}}
\end{aligned}
\end{equation}

This is in fact our ``global" cost function. A ``local" cost function is defined as such:
\begin{equation}
\begin{aligned}
    &\hat{C}_L =\\ 
    &\frac{1}{2} - \frac{1}{2n} \frac{\sum_{l, l', j} c_l c_{l'} \bra{0} V(\boldsymbol{\theta})^\dagger A^\dagger_l U Z_j U^\dagger A_{l'} V(\boldsymbol{\theta}) \ket{0}}{ \sum_{l, l'} c_lc_{l'}\bra{0}V(\boldsymbol{\theta})^{\dagger}A_l^{\dagger}A_{l'}V(\boldsymbol{\theta})\ket{0}}
\end{aligned}
\end{equation}
where the $\ket{0}\bra{0}$ from $\hat{C}_G$ has been replaced by the positive operator
\begin{align}
    P = \frac{1}{2} + \frac{1}{2n}\sum_{j=0}^{n-1} Z_j.
\end{align}

The local cost function is shown to avoid barren plateau problems that appear in the global cost function \cite{vqlspap}. Both of these cost functions were tested while, but due to the severe barren plateau problems that appear for the global cost function, all results shown are from utilizing the local cost function.  

The distance between the quantum state $\ket{\psi(\boldsymbol{\theta})}$ and the real solution (as represented by a proportional quantum state, $\ket{x}$), is denoted by $\epsilon$, which is defined by
\begin{align}
    \epsilon = \frac{1}{2}Tr| \ket{x}\bra{x} - \ket{\psi}\bra{\psi}|.
\end{align}
We have inherited this definition as well as the operational relation between the local and global cost functions that is introduced in \cite{vqlspap}, which includes the following bounds:

\begin{align}
    \hat{C}_G \geq \frac{\epsilon^2}{\kappa^2}, C_G \geq \frac{\epsilon^2}{\kappa^2}, \hat{C}_L \geq \frac{\epsilon^2}{n \kappa^2}, C_L \geq \frac{\epsilon^2}{n\kappa^2}, 
\end{align}
where $\kappa$ is the condition number of the matrix, and $n$ is the number of qubits defining the size of the matrix. We use these bounds in our numerical simulations, where the optimization routine attempts to reduce the cost function below these thresholds in order to reach a desired $\epsilon$ error (set by the user), given the $n$ and $\kappa$ attributed to the matrix.

For the sake of simplicity, we took $\ket{b} = H^{\otimes n} \ket{0}$. 

\subsubsection{Quantum Circuit Evaluations per Local Cost Function Evaluation}

Due to the nature of the cost function definitions including multiple summations over expectation values, there is a need to quantify exactly how many quantum circuit evaluations are required to complete a single cost function evaluation. What is referred to as a "quantum circuit evaluation" is a job running 1e6 shots of the same unique quantum circuit in order to evaluate the expectation value of some term in the cost function. The number 1e6 shots was used because it is the maximum number of shots one could run on the IonQ quantum computers. In most literature, number of iterations are usually presented in the main results (for example, in \cite{vqlspap, delft}, etc). However, the number of quantum circuits required to reach a solution is a much more practically important convergence metric. To illustrate this point, one could imagine a cost function that has a blowup in the number of quantum circuits required to evaluate the cost function with respect to system size. A contrived example of this is if one were to use the Pauli decomposition of the DPEM, which would require the number of circuit evaluations for a cost function evaluation to scale exponentially with respect to system size. Most of the time, the dependence on system size is linear at worst, but the example still demonstrates the importance of communicating exactly how one's cost function computational scales. 

As such, we now will show the number of quantum circuits required to run the local cost function, since the global cost function is not the focus of this paper. In order to reduce the number of quantum circuit required to evaluate the local cost function, one can use the following equivalences:

\begin{align}
    l = l' \Longrightarrow \bra{0}V(\boldsymbol{\theta})^{\dagger}A_l^{\dagger}A_{l'}V(\boldsymbol{\theta})\ket{0} = 1
\end{align}
\begin{equation}
\begin{aligned}
    \bra{0}V(\boldsymbol{\theta})^{\dagger}&A_l^{\dagger}A_{l'}V(\boldsymbol{\theta})\ket{0} =\\
    &\bra{0}V(\boldsymbol{\theta})^{\dagger}A_{l'}^{\dagger}A_{l}V(\boldsymbol{\theta})\ket{0}
\end{aligned}
\end{equation}
\begin{equation}
\begin{aligned}
    \bra{0} V(\boldsymbol{\theta})^\dagger &A^\dagger_l U Z_j U^\dagger A_{l'} V(\boldsymbol{\theta}) \ket{0} =\\
    &\bra{0} V(\boldsymbol{\theta})^\dagger A^\dagger_{l'} U Z_j U^\dagger A_{l} V(\boldsymbol{\theta}) \ket{0}
\end{aligned}
\end{equation}

After doing these reductions, one finds that the number of quantum circuits required to evaluate the local cost function is 

\begin{align}
    N_{q} &= n  \sum_{i=1}^c i + \sum_{i=1}^{c-1} i\\
    &= \frac{1}{2}c[n(c+1)+c-1],
\end{align}
where $c$ is the number of sub-matrices in the decomposition of $A$. For the HED, which all of our numerical results use (to avoid the exponential blowup of the Pauli decomposition), $c=4$, so we see that $N_q$ has a mild linear dependence on system size. 
\begin{align}
    N_q &= 10n + 6\\
    &= O(n)
\end{align}

One might now understand how $c = O(2^n)$ would cause an exponential blowup of the time required to evaluate a single cost function. While we know $n$ is still exponentially related to system size, this would result in the cancellation of any speedups we might hope to achieve with quantum computing. We will later show how the number of quantum circuits required per cost function evaluation affects the time-to-solution on real quantum computers. 

\textit{Brief Note on Parallelization:} One might notice that these quantum circuits may be parallelized to improve predicted time-to-solutions metrics that we will show below (following a method similar to \cite{parallel}). However, according to IonQ Application Scientist Dr. Daiwei Zhu \cite{daiwei}, doing so would introduce more noise into the circuits, making it much more difficult to train the ansatz. It is best to not operate under the parallel assumption at all in this case, so in this paper, we make no mention of a speedup introduced with parallelization.
\begin{figure*}[ht]
    \centering
    \subfigure[]{\includegraphics[height=0.175\textheight]{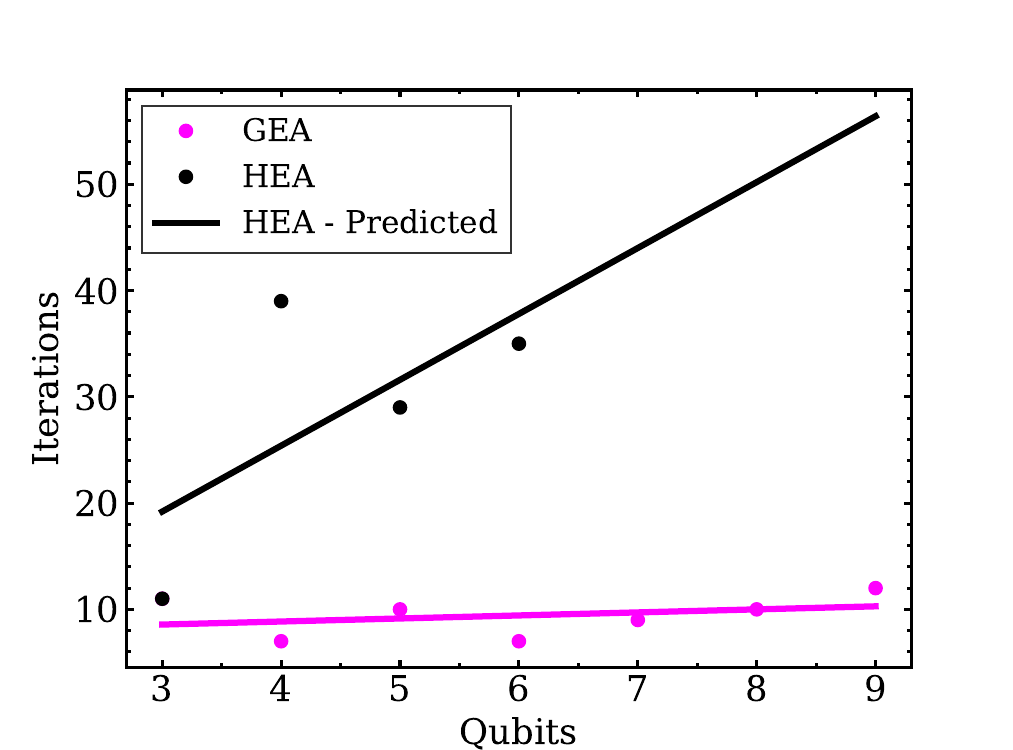}}
    \subfigure[]{\includegraphics[height=0.175\textheight]{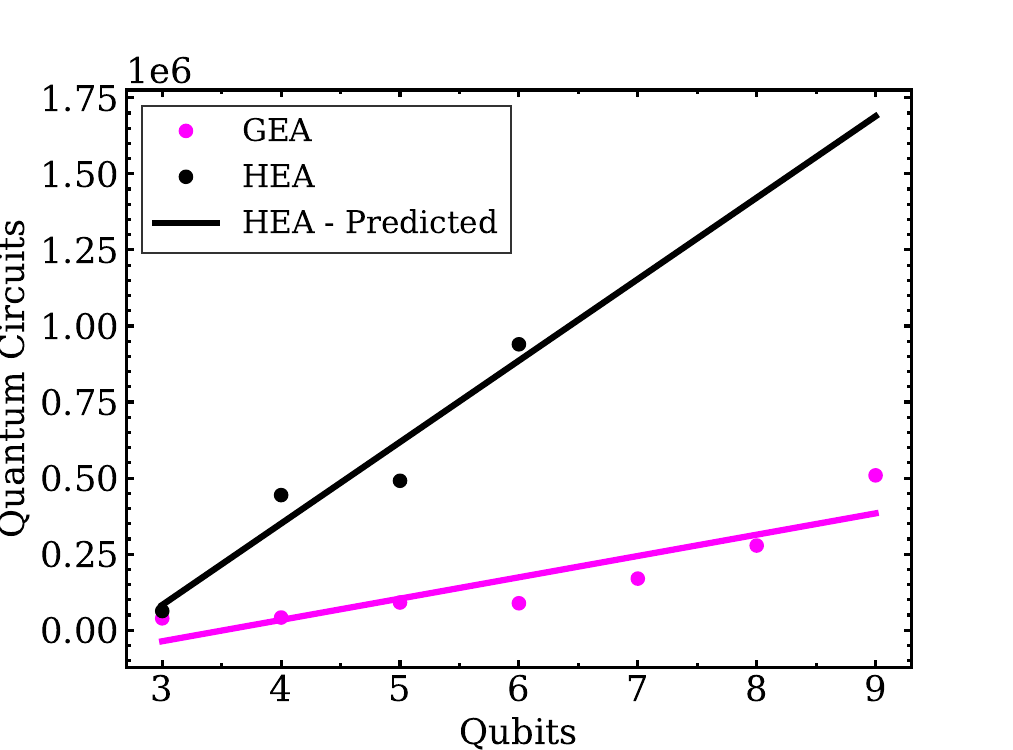}}
    \subfigure[]{\includegraphics[height=0.175\textheight]{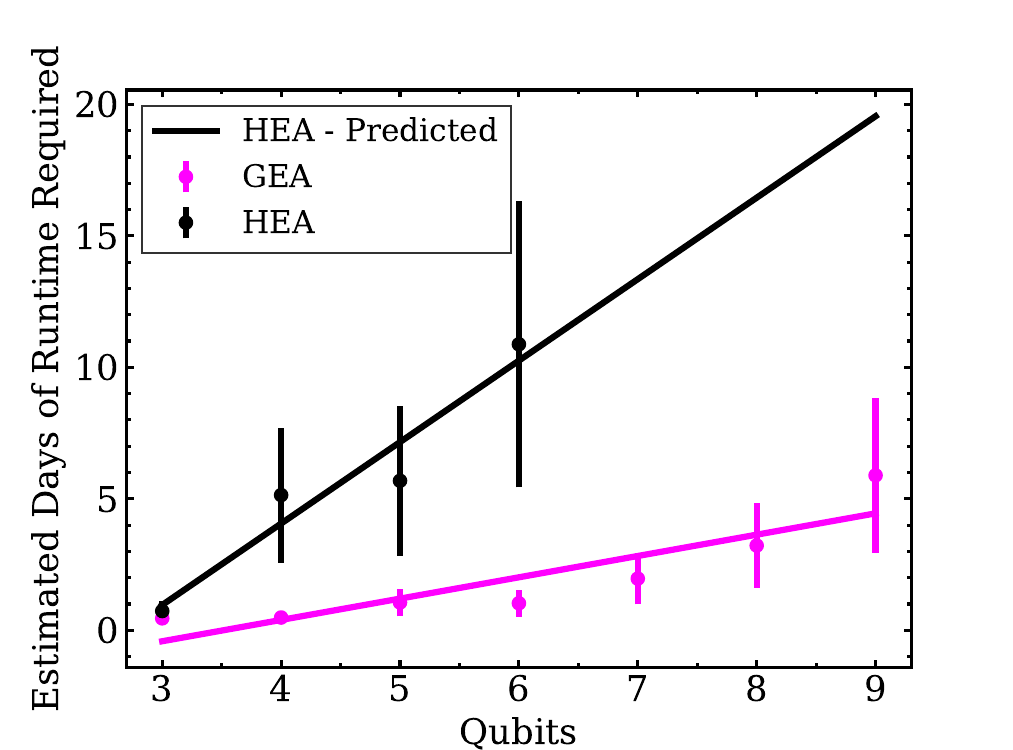}}
    \caption{Iterations (a), quantum circuit evaluations (defined as a job running 1e6 shots of the same quantum circuit) (b), and estimated days (c) versus qubits defining the system size for both HEA and GEA runs. Here we see that the GEA has a much favorable scaling of all metrics of time-to-solutions. While all 3-9 qubit systems were able to complete without time-out, the HEA would only perform up to 6 qubit systems. The error-bars on (c) are because IonQ experts have said that an average quantum circuit evaluation running on an IonQ quantum computer would take $1\pm 0.5$ minutes \cite{daiwei, ionqrun}.}
    \label{fig:0_01_metrics}
\end{figure*}
\section{Numerical Results}

In the following numerical results, we utilized the Pennylane library for quantum circuit representation. Since we found that the Pauli decomposition required an exponential scaling of the sub-matrices, we used the HED of the Poisson in all cases. System sizes were only considered from 3 to 9 qubit systems. We noticed that, although both the global and local cost functions could successfully minimize and output the correct solution vector, for higher qubit systems, the global cost function often encounters barren plateau problems. For this reason, all our numerical simulations used the local cost function in order to give both ansatz the chance to converge to a solution. Also, we tested two different variances in the randomized weight initialization, $q_{\Delta} = 0.01$ and $q_{\Delta} = 0.1$. We found that the HEA had the most success with $q_{\Delta} = 0.01$, so we use them as an example in our results section. However, all remaining simulations done with $q_{\Delta} = 0.1$ are shown in Appendix \ref{appd:b} and can be used to draw the same conclusions. 

\subsection{Practical Constraints}

The goal of this paper is to provide \textit{truly} near-term quantum algorithms that could be used to solve complex systems of equations, such as the DPEM, within the next 5-10 years. As such, we believe that the practical constraints (that will be explained below) we placed on our numerical simulations were justified. 

Since we used Pennylane's fastest numerical quantum simulator, the Lightning Qubit, our tests were done at a much faster pace than any state-of-the-art quantum computer (which would be far slower due to the sampling and validation requirements completed on the back-end). For this reason, we imposed a practical limitation on our numerical testing, namely that we would not use super-computers or high-performance GPUs to further speed up the computational process. All the simulations were run on a Google Colab notebook, which had a time-out period of 12 hours. If a run of the simulator did not complete sufficient iterations before the 12-hour mark (i.e., getting past more than one or two iterations), we removed those results from the data. In reality, simulations that we see take near 8-10 hours on the Pennylane Lightning Qubit simulator would take upwards of days on a real quantum computer. At this order of magnitude, allowing the use of super-computers or high-performance GPUs to speed-up the simulations would be pointless; even if these results worked, the practical implementation of such enormous numbers of quantum circuit evaluations would take unreasonably long to run on quantum computers scientists have access to within the next 5-10 years.  

To better understand this problem, we have shown the time-to-solution metrics as both number of iterations, cost function evaluations, and estimated days required to converge to a solution with respect to system size. As per IonQ's website \cite{ionqrun} and the advice of IonQ Application Scientist Dr. Daiwei Zhu, each quantum circuit evaluation takes $1 \pm 0.5$ minutes to run. Using this metric, we calculated the estimated number of days it would require to complete one run of the VQLS for solving the DPEM. We also compared these numbers to the less daunting iteration counts per run with respect to system size in Figure \ref{fig:0_01_metrics}. It is worth noting that these estimated times assume that the result of each quantum circuit evaluation offers a sufficient precision that is required to minimize below the convergence threshold. More on this is examined in Section \ref{disc}.

\subsection{Comparison of the HEA and GEA}

We tested the difference in convergence metrics between using the HEA and the GEA with numerical, state-vector simulations. In Figure \ref{fig:0_01_metrics}(a), we see the iterations scaling is linear for both the HEA and GEA, but the GEA has a far better scaling compared to the predicted scaling of the HEA. If we only examined the iteration scaling, however, we would miss out on a more interesting metrics. For example, in Figure \ref{fig:0_01_metrics}(b), we can see that what we initially thought was a benign linear scaling with respect to qubits turns out to have significant consequences for the number of quantum circuit evalutations required to converge to a solution. For the HEA, the simulated and predicted number is in the millions, with a large-coefficient linear scaling, while the GEA remains below 0.5e6 quantum circuit evaluations and has a much better coefficient for its linear scaling. Finally, when we examine the estimated days of runtime required for solving each system, we can see again the significant consequences of a bad linear coefficient. While the GEA is constrained to no more than 5 days for system sizes $n=3$ to $n=9$, the predicted scaling of the HEA exposes how unreasonably long the time-to-solutions would be if one were to run them on real quantum computers. Ideally, our algorithms would be best if the number of quantum circuits required to run was kept at a minimum, since there are many time-consuming components of running a quantum circuit on a real quantum computer that cannot be represented in quantum simulators like Pennylane. 

\subsection{Scaling of $\kappa$}

In the special case of the DPEM, the condition number scales exponentially with the number of qubits, as can be seen in Figure \ref{fig:cond_vs_qubits} (note that this simply means that its condition number scales linearly with respect to system size). As shown in \cite{Harrow2009}, linear scaling of iterations with respect to condition number is optimal in the general case. Since condition number is exponentially related to qubits, and the iterations scale linearly with with respect to qubits for both the HEA and GEA, both methods are sub-linear, indicating at least optimally in both cases. This sub-linear scaling with respect to system size is found also in \cite{vqlspap}, and is thought to be due to the special case of the problem being examined.  

\begin{Figure}
    \centering
    \includegraphics[height=0.25\textheight]{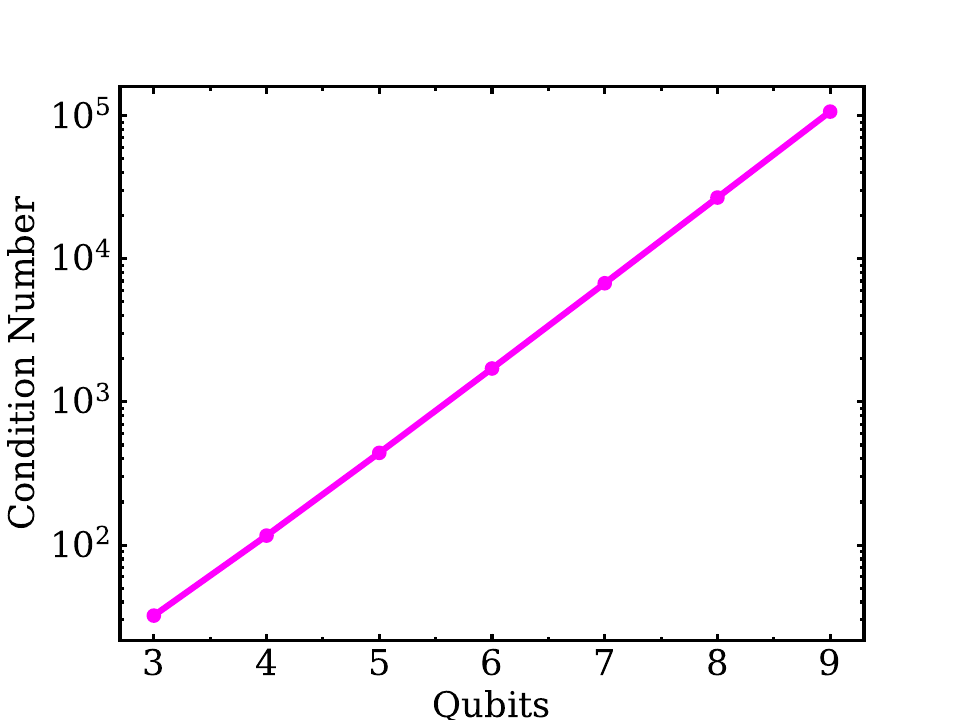}
    \captionof{figure}{Condition number versus qubits defining solution vector. Note that the y-axis is logarithmic scaling, meaning that the DPEM scales exponentially in condition number with respect to qubits. However, this does not mean it scales exponentially with respect to system size; since qubits are related to system size exponentially, condition number scales linearly with respect to system size.}
    \label{fig:cond_vs_qubits}
\end{Figure}

\begin{figure*}[ht]
    \centering
    \subfigure[]{\includegraphics[width=0.43\textwidth]{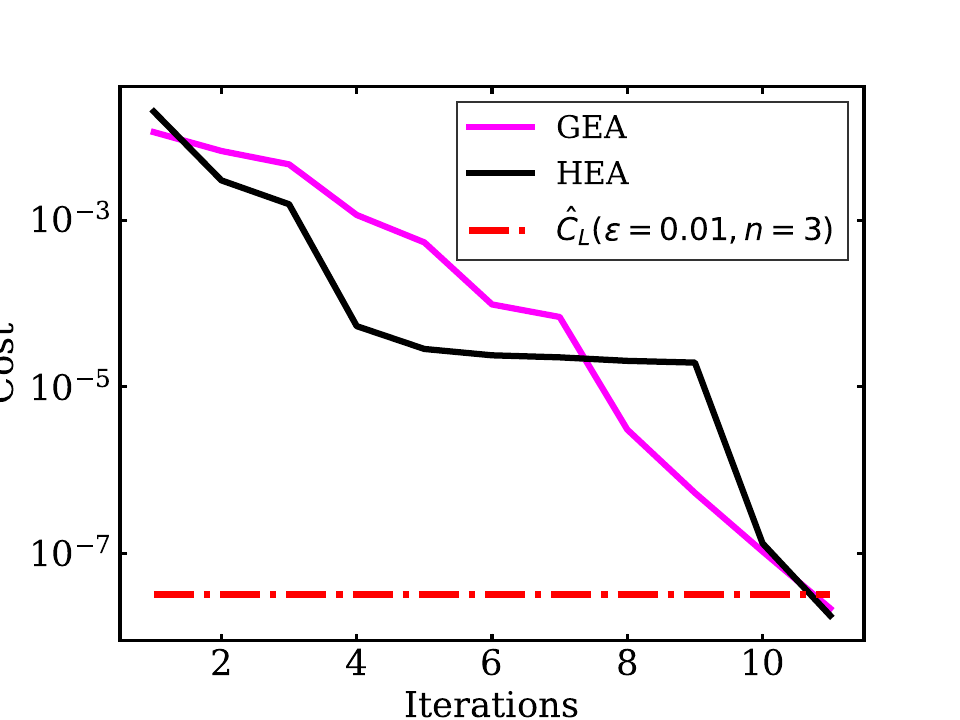}}
    \subfigure[]{\includegraphics[width=0.43\textwidth]{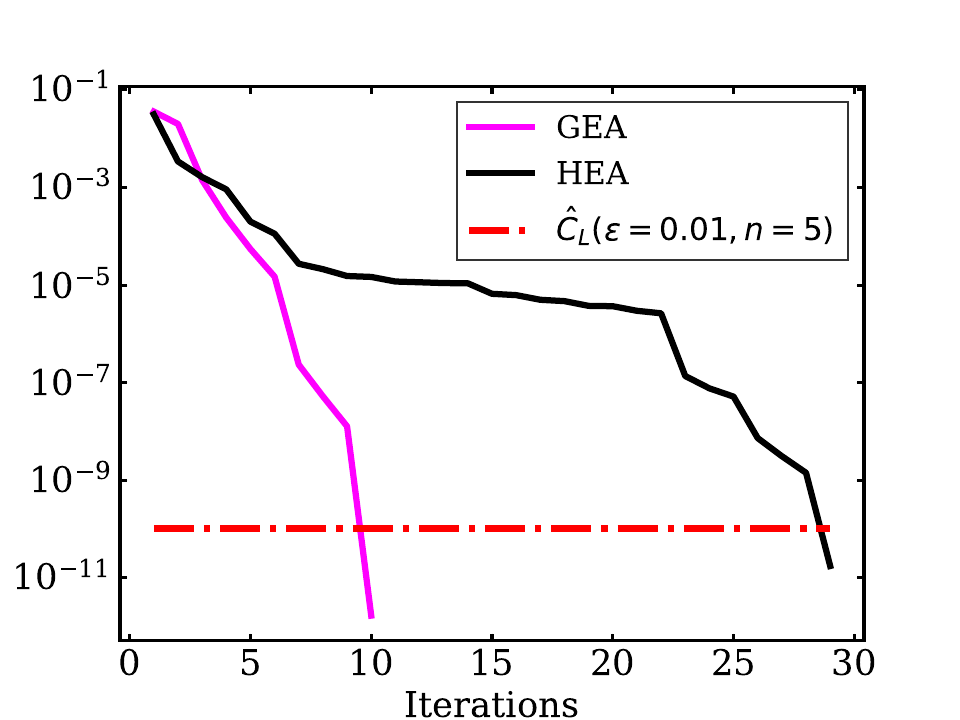}}
    \subfigure[]{\includegraphics[width=0.43\textwidth]{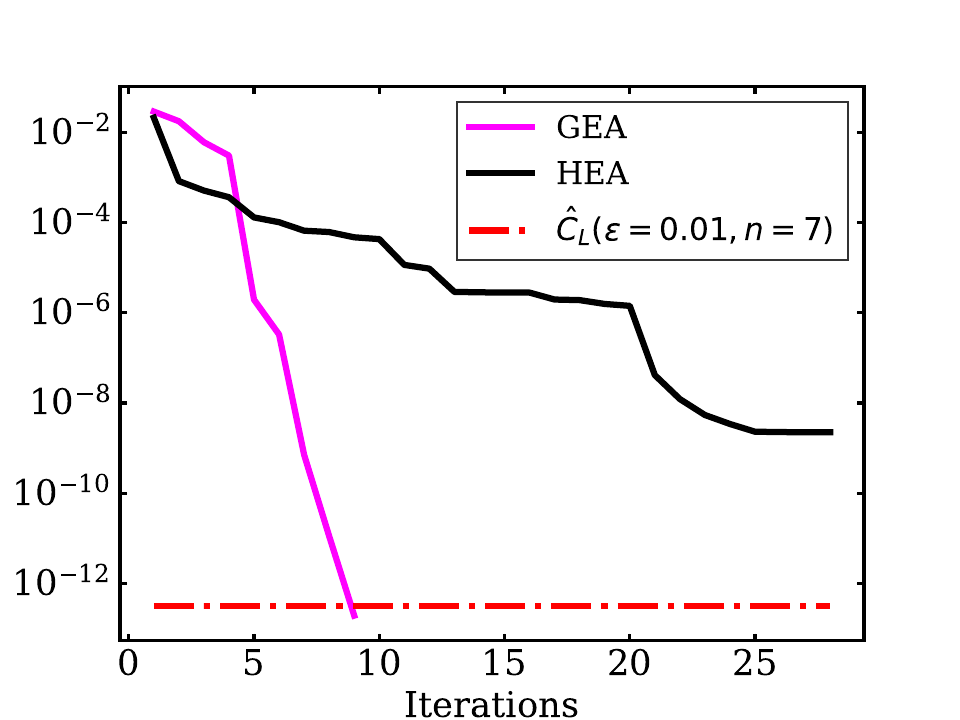}}
    \subfigure[]{\includegraphics[width=0.43\textwidth]{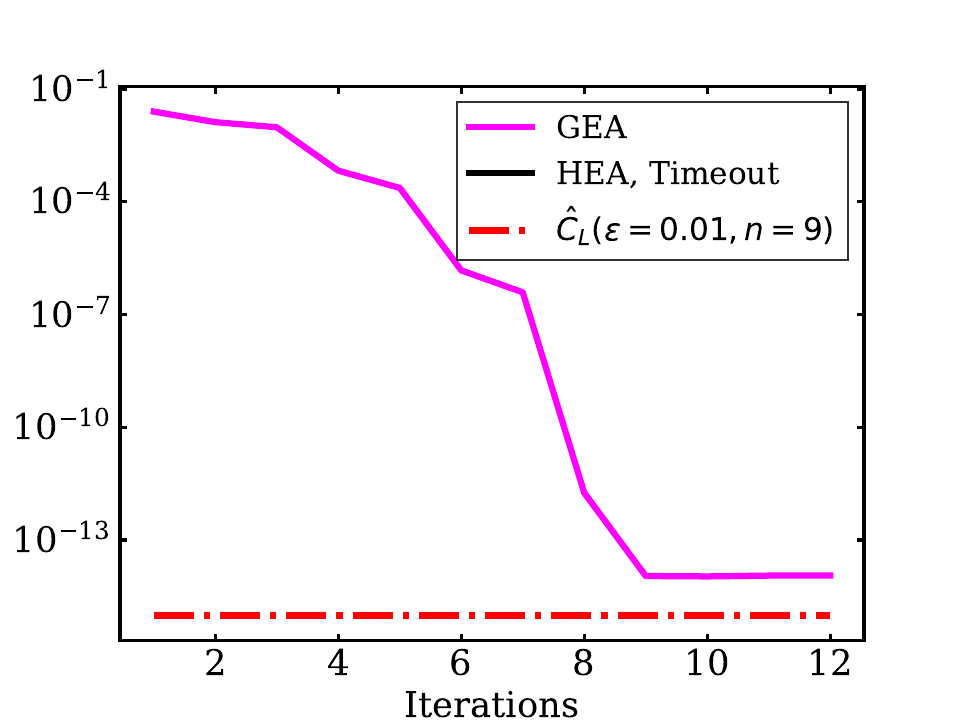}}
    \caption{Results of numerical simulations with both the HEA and GEA, using the local-cost function. The target $\epsilon = 0.01$ for all runs, although at higher qubit systems, it is not always possible to reach that precision. One can clearly see that the GEA converges much faster than the HEA, which sometimes fails to converge at all or fails to even complete a sufficient number of iterations within the time-out period of 12 hours. }
    \label{fig:0_01_qubits}
\end{figure*}

\section{Discussion}\label{disc}
The results of our work show that crafting VQAs under the assumption of all-to-all connectivity of qubits, utilizing their high-entanglement capabilities for both ansatz and encoding on quantum computers, may be a promising path forward towards the goal of near-term, practical quantum algorithms. Our paper proposes a novel ansatz, the GEA, which offers many computational improvements compared to most ansatz derived in the literature. With the entanglement of all qubits with each other at each layer, the expressibility of our ansatz is greatly increased without the need for more rotation gates. This makes it easier for an optimizer to converge to a solution, since the parameter space has far less dimensions compared to typical hardware-efficient ansatz. We saw that, while both the GEA and HEA had linear scaling of parameter space with respect to qubits, the GEA had a far better linear coefficient, as seen in Figure \ref{fig:params_vs_qubits}. The numerical simulations we conducted confirms our conjecture that the reduction in parameter space enables the optimizer to converge far quicker with the GEA, as shown in Figure \ref{fig:0_01_qubits}. This means that the time-to-solution would be considerably quicker when running the algorithm with a GEA on a real quantum computer compared to running it with an HEA (see Figure \ref{fig:0_01_metrics}). 

Additionally, our example of the HED for the DPEM shows that using high-entanglement methods to represent matrices or other, more complicated structures on a quantum computer may be the key to unlocking better quantum algorithms. This paper showed that the HED offers a constant-scaling decomposition of the DPEM compared to the Pauli decomposition, which is $O(2^n)$, while remaining under a reasonable circuit depth scaling of $O(n^2)$. Note that some other papers have been written on reducing the decomposition size of the DPEM, but our paper offers unique insights and better decompositions than these other papers. \cite{pois1} proposes a decomposition that has $O(logn)$ sub-matrices, while our decomposition is $O(1)$. \cite{pois2} proposes another $O(1)$ decomposition, but the authors do not discuss the circuit depth requirements of their decomposition. For example, the shift operator is mentioned, but the consequences on the required circuit depth for cost function evaluations and time-to-solution metrics are not examined closely. Our paper indeed examines the bounds on circuit depth that come as a trade-off when implementing our $O(1)$ decompositions and the resultant effect on the practicality of this algorithm. 

These are only a few of the many potential ideas that may unlock the capabilities of high-entanglement quantum computers such as the IonQ Aria. Other applications of high-entanglement capabilities include the usage of GEA in other quantum machine learning tasks as outlined in \cite{qml, qml2} or reducing the number of ancilla qubits required for decomposing Quantum Phase Estimation (QPE) routines into native gates \cite{qpe} (such reductions are examined in papers such as \cite{qpe2} without the use of high-entanglement). A similar procedure may be done with routines such as Quantum Amplitude Estimation \cite{qpa} or Quantum Mean Estimation \cite{qme}. Future work is still necessary to be done in order to examine more closely other ways in which high-entanglement capabilities may be used to speed-up quantum computing algorithms, whether through a reduction of circuit depth, simplification of encoding quantum states, or shortening convergence times of other heuristic algorithms. 

\subsection{Acknowledgements}
The authors of this paper would like to thank the team at IonQ and QLab that gave us the resources to support our work, both for computing capabilities and knowledge from their staff. We would specifically like to thank Dr. Franz J. Klein and Dr. Daiwei Zhu for the invaluable guidance given to the authors about this project. We would also like to thank the Johns Hopkins Undergraduate Society for Applied Math for inviting us to present our work at the 2024 Mid-Atlantic Research Exchange (MATRX) in both a lecture and poster-presentation. 

\subsection{Statement of Funding}

The research conducted that culminated into the contents of this paper was funded by the QLab/IonQ Seed Grant for \$50k, which was granted to Dr. James Baeder (PI) and  Rahul Babu Koneru (Co-PI) for use in the project titled ``Variational Quantum Linear Solver for Computational Fluid Dynamics". 
\end{multicols}

\newpage
\begin{multicols}{2}
    \bibliographystyle{unsrt}
    \bibliography{ref}  
\end{multicols}  

\newpage
\section*{Appendix}
\appendix\section{6-qubit Circuit Representations}\label{appd:a}
\setcounter{figure}{0}
\renewcommand{\figurename}{A.}

\begin{figure*}[h]
    \centering
    \subfigure[]{\includegraphics[scale=0.25]{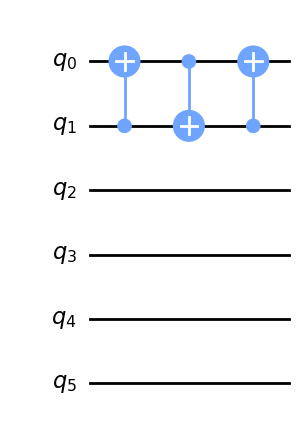}}
    \subfigure[]{\includegraphics[scale=0.25]{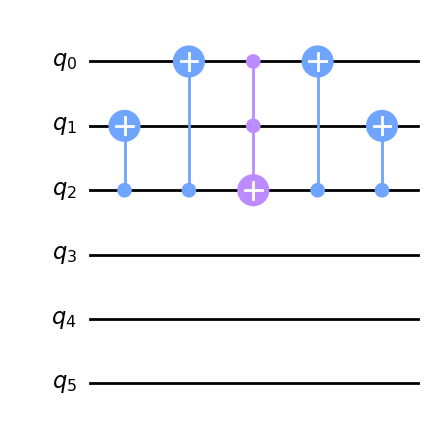}}
    \subfigure[]{\includegraphics[scale=0.25]{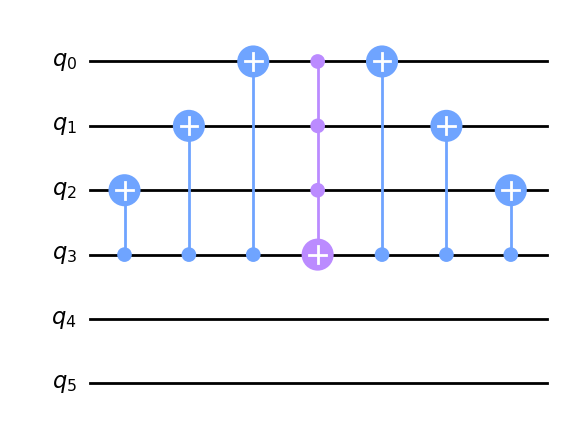}}
    \subfigure[]{\includegraphics[scale=0.25]{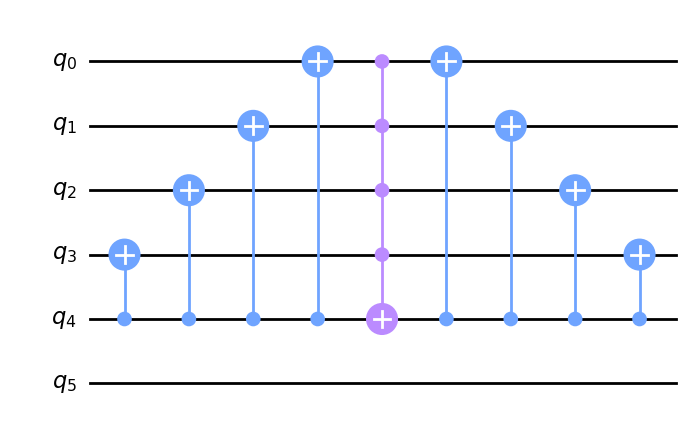}}
    \subfigure[]{\includegraphics[scale=0.25]{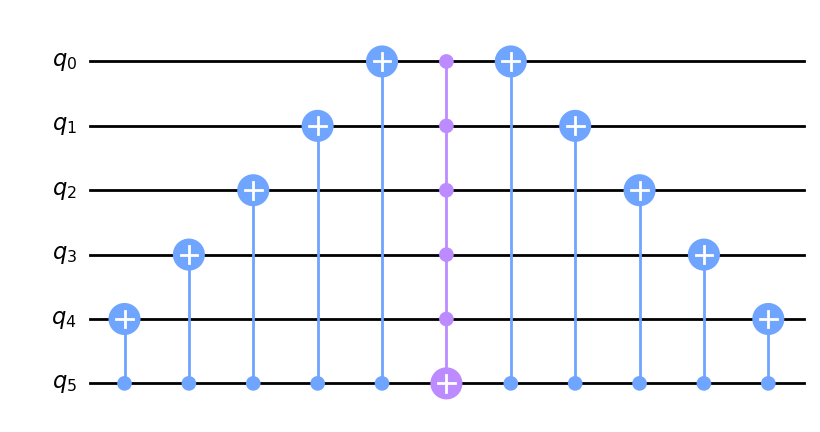}}
    \caption{$C_1$, $C_2$, ..., and $C_5$ from left to right, which will be used to construct the circuit for the $L_2$ matrix.}
    \label{fig:Cgates}
\end{figure*}
\begin{figure*}[h]
    \centering
    \includegraphics[scale=0.3]{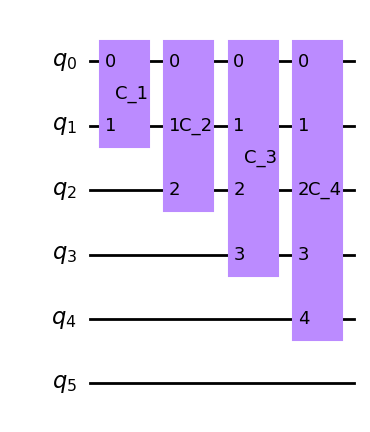}
    \caption{The final circuit that represents the application of the $L_2$ matrix to the ansatz.}
    \label{fig:Cfinal}
\end{figure*}
\newpage
\begin{figure}[h]
    \centering
    \includegraphics[scale=0.3]{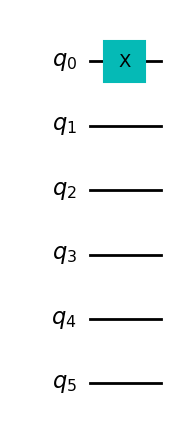}
    \caption{Quantum circuit representing $L_1$}
    \label{fig:L1}
\end{figure}
\begin{figure}[h]
    \centering
    \includegraphics[scale=0.3]{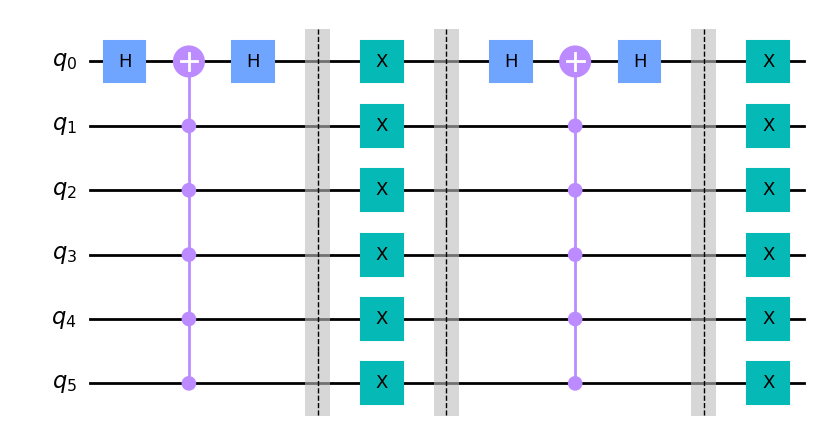}
    \caption{Quantum circuit representing $L_3$}
    \label{fig:L3}
\end{figure}

\newpage

\section{Remaining Numerical Simulations}\label{appd:b}
\setcounter{figure}{0}
\renewcommand{\figurename}{B.}
\begin{figure}[ht]
    \centering
    \subfigure[]{\includegraphics[width=0.255\textwidth]{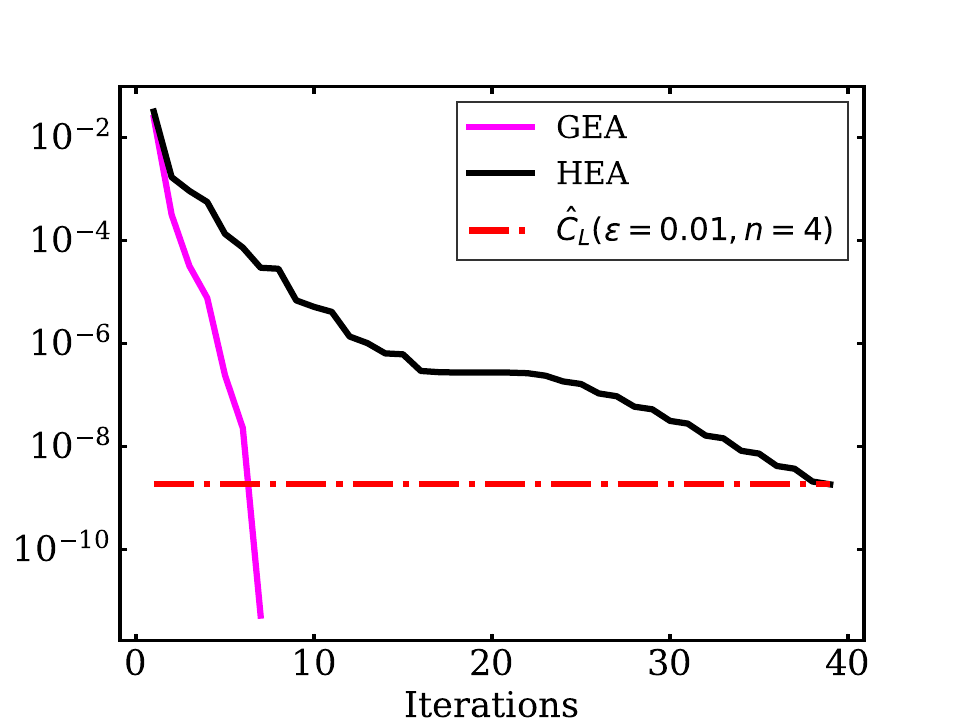}}
    \subfigure[]{\includegraphics[width=0.255\textwidth]{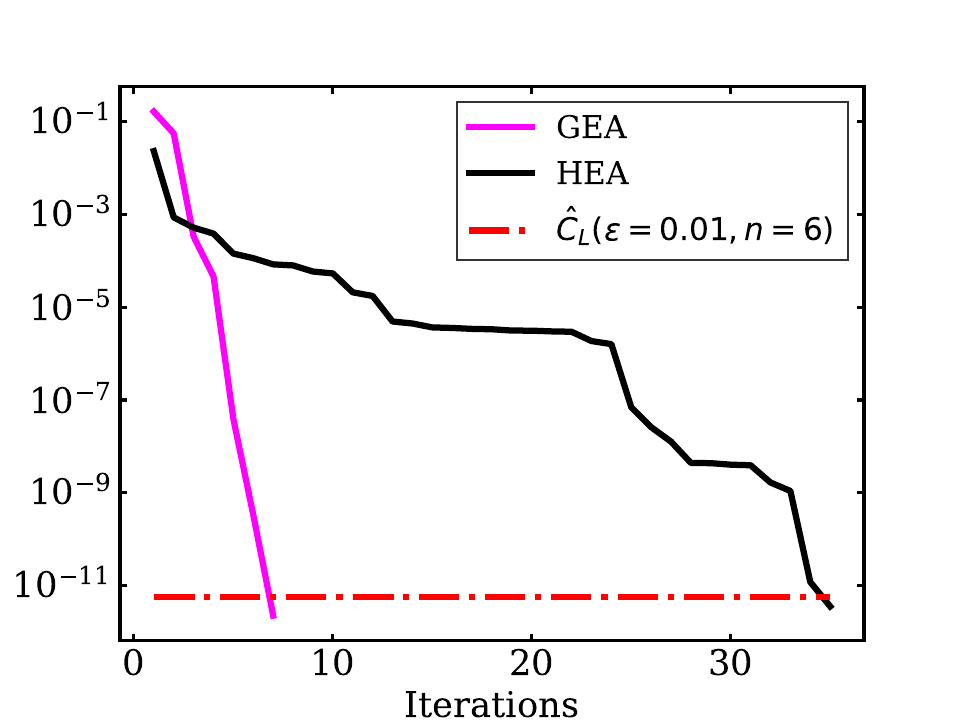}}
    \subfigure[]{\includegraphics[width=0.255\textwidth]{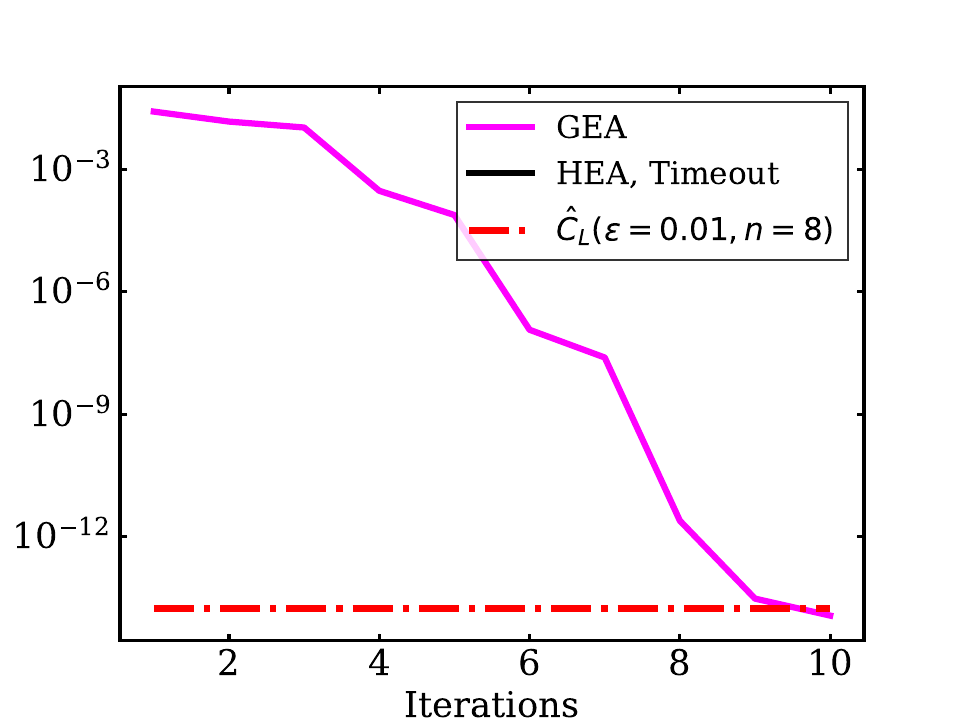}}
    \caption{Remaining runs of $q_{\Delta} = 0.01$ that were not shown in the figures above. }
    \label{fig:0_01_qubits_rem}
\end{figure}
\begin{figure}[ht]
    \centering
    \subfigure[]{\includegraphics[width=0.255\textwidth]{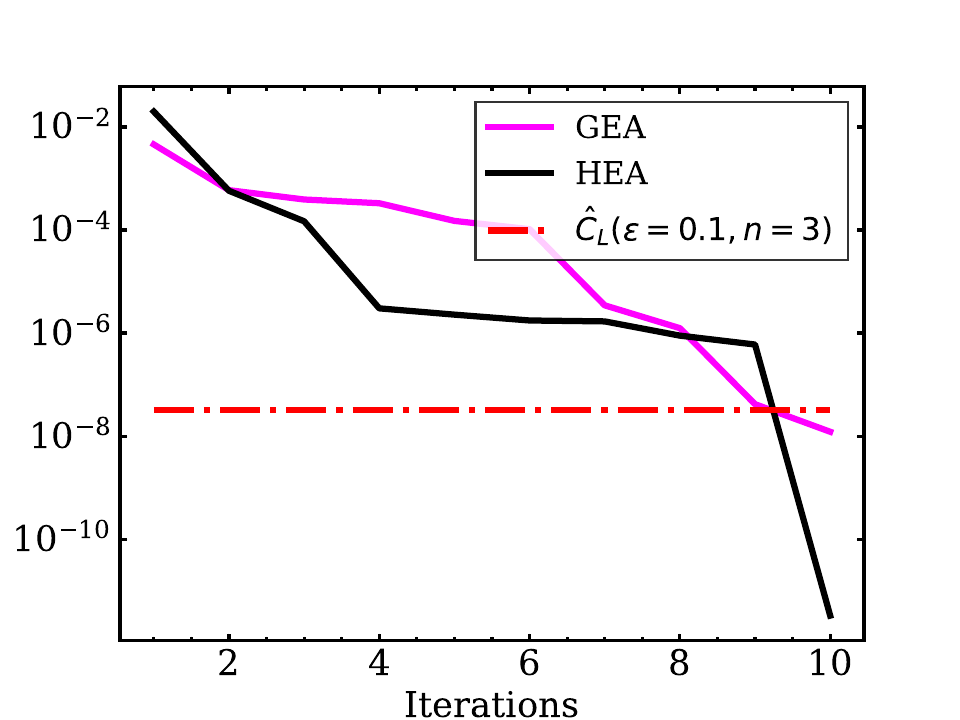}}
    \subfigure[]{\includegraphics[width=0.255\textwidth]{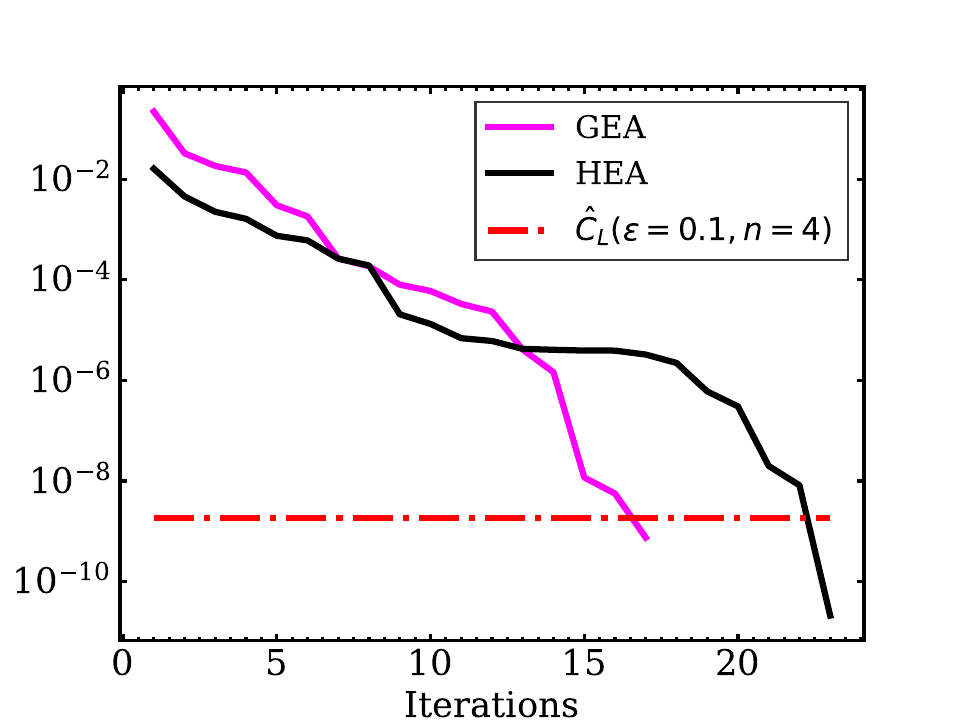}}
    \subfigure[]{\includegraphics[width=0.255\textwidth]{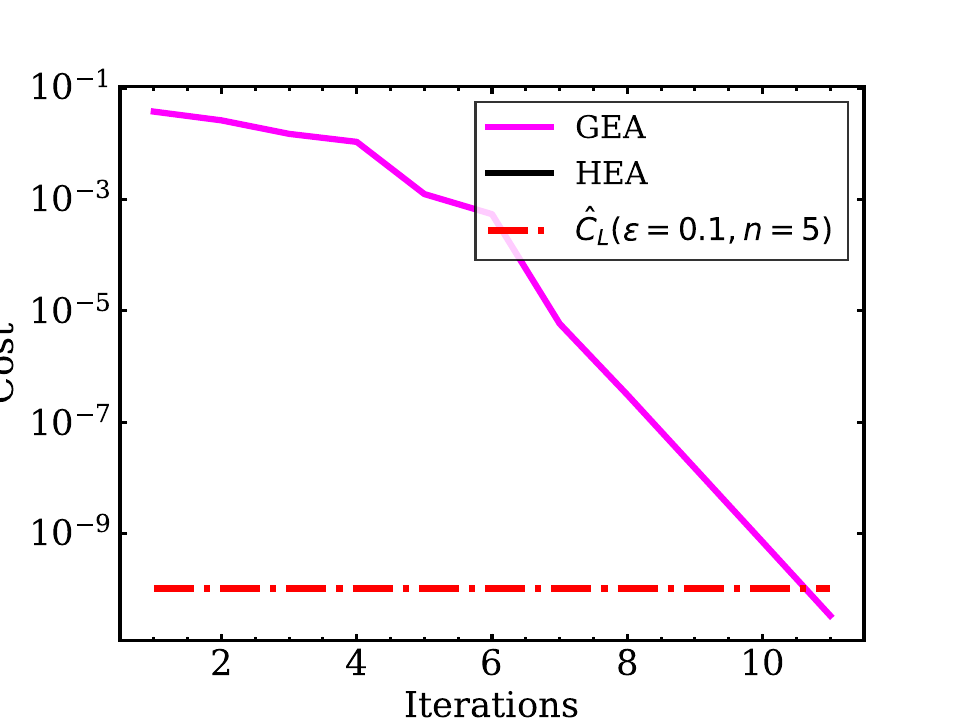}}
    \subfigure[]{\includegraphics[width=0.255\textwidth]{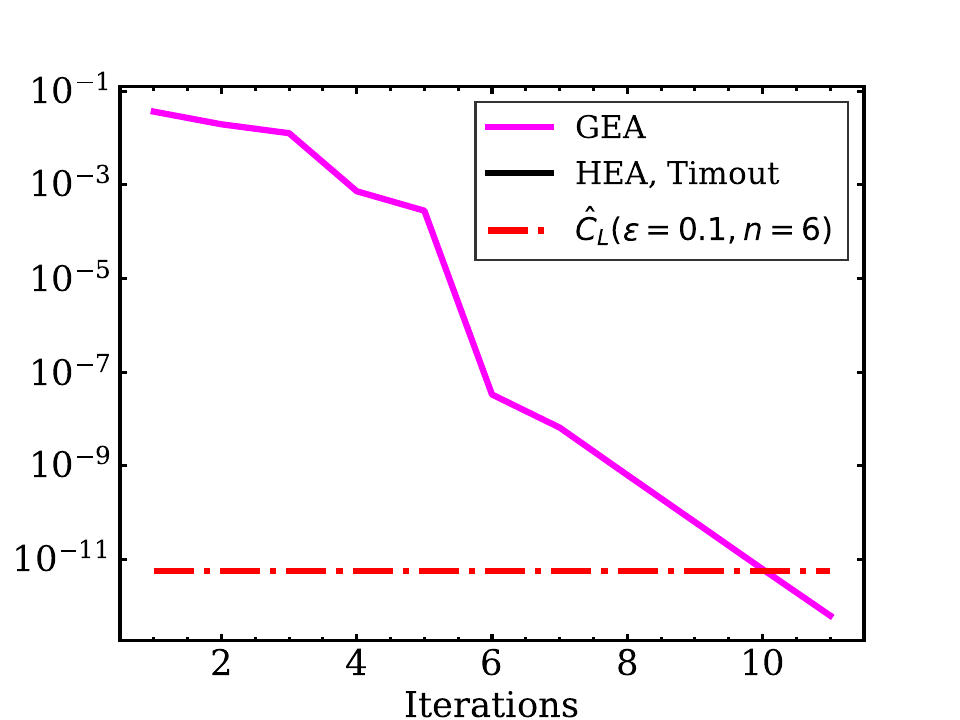}}
    \subfigure[]{\includegraphics[width=0.255\textwidth]{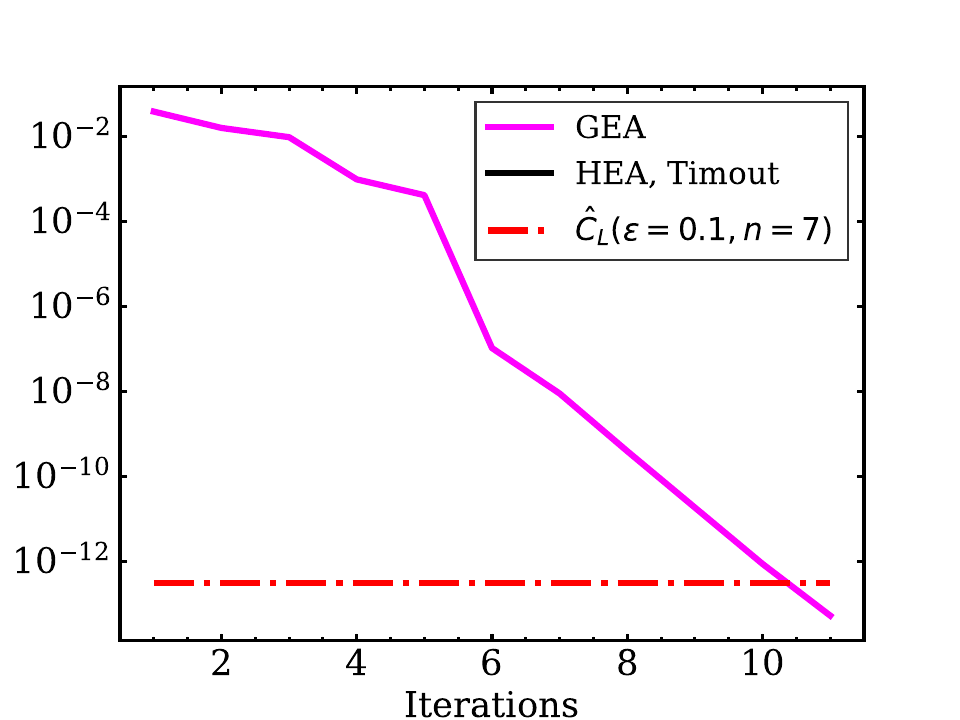}}
    \subfigure[]{\includegraphics[width=0.255\textwidth]{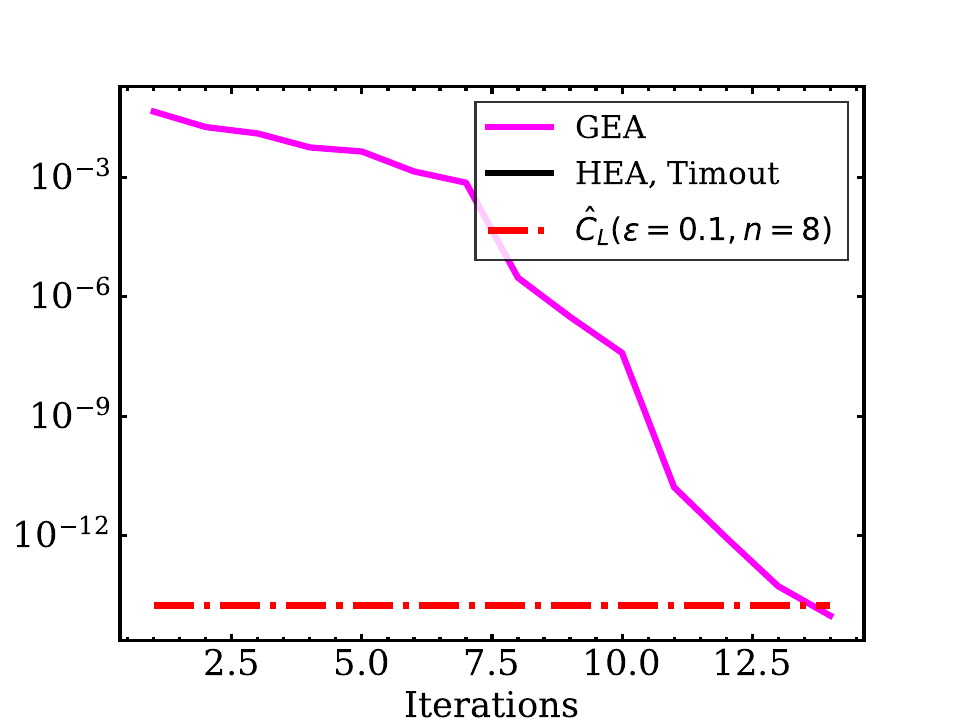}}
    \subfigure[]{\includegraphics[width=0.255\textwidth]{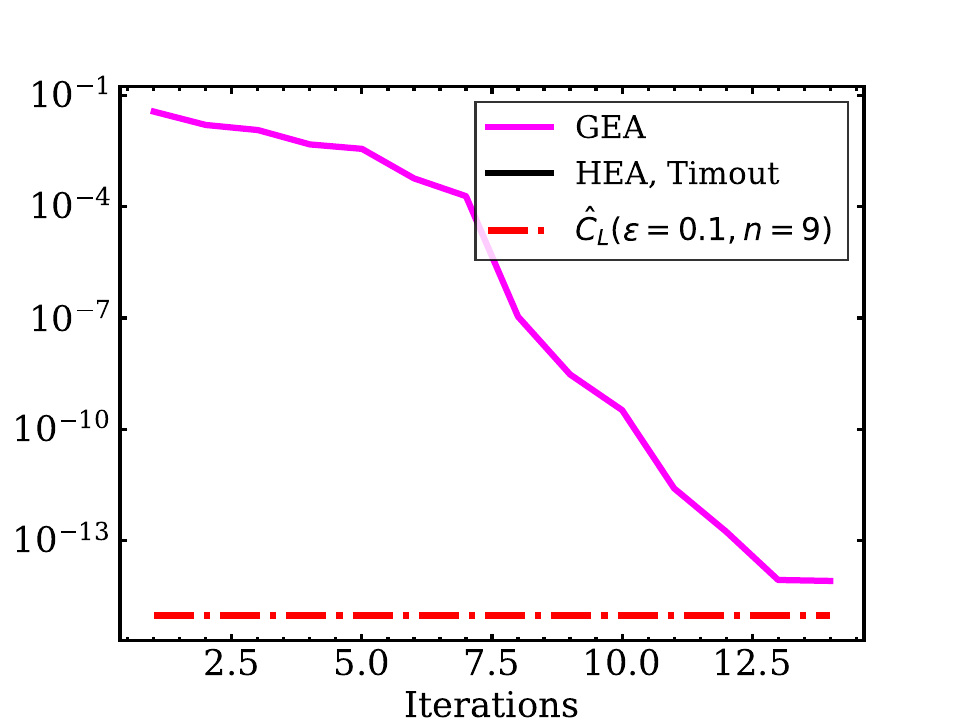}}
    \caption{Results of numerical simulations with $q_{\Delta} = 0.1$. Here we see that the simulation times out after running simulations of system sizes larger than $n=4$, whereas the GEA does not have the same timeout problems. The simulations timed-out after iterations took longer than 12 hours to run. Simulations were run until they reached the precision metric defined in \cite{vqlspap}.}
    \label{fig:0_1_qubits}
\end{figure}

\end{document}